\documentclass[prl,twocolumn,showpacs,amssymb,aps]{revtex4-1}
\usepackage{dcolumn}
\usepackage{bm}
\usepackage{graphicx}
\usepackage{epstopdf}
\usepackage{subfig}
\usepackage{color}
\usepackage{amsmath}
\usepackage{tikz-cd}
\usepackage[colorlinks=true]{hyperref}
\begin{document}
	\title{Rogue wave multiplets in the complex KdV equation}
	\author{M. Crabb and N. Akhmediev}
	\affiliation{Optical Sciences Group, Department of Theoretical Physics, Research School of Physics, The Australian National University, Canberra, ACT, 2600, Australia}
	\begin{abstract} 
	We present a multi-parameter family of rational solutions to the complex Korteweg--de Vries (KdV) equations. This family of solutions includes particular cases with high-amplitude peaks at the centre, as well as a multitude of cases in which high-order rogue waves are partially split into lower-order fundamental components. We present an empirically-found symmetry which introduces a parameter controlling the splitting of the rogue wave components into multi-peak solutions, and allows for nonsingular solutions at higher order.
	\end{abstract}
	\maketitle
	\date{today}
\section{Introduction}
Rogue waves are known to exist on deep ocean surfaces \cite{Kharif,Osborne}, within water in the form of internal rogue waves \cite{Grimshaw,Alford}, in optical fibres in supercontinuum generation \cite{Solli,Dudley}, in the vacuum in the form of quantum fluctuations \cite{Chekhova} and even in the theory of gravitational waves \cite{Bayindir}. Their universality has been confirmed by water tank experiments \cite{Chabchoub}, in quadratic nonlinear crystals \cite{Schiek}, and, most strikingly, in our encounters with extreme natural phenomena \cite{Nikolkina}. 
The most common approach to the description of rogue waves is using the exact solutions of integrable evolution systems such as three-wave interaction \cite{Baronio}, nonlinear Schr\"odinger (NLS) \cite{Gaillard,Trulsen}, Kadomtsev-Petviashvili \cite{Dubard,Kodama}, and Davey-Stewartson \cite{Ohta} equations, among others \cite{Zhang}. Rational Peregrine-like solutions of these equations provide a good approximation to describing rogue wave formation in a variety of physical situations \cite{Shrira,Onorato}.\\\indent
The real Korteweg-de Vries (KdV) equation \cite{Korteweg} is the basis of the most common tool for the (1+1)-dimensional modelling of shallow water waves, which has been in use since the work of Boussinesq \cite{Boussinesq}. Numerical modelling done by Zabusky and Kruskal revealed the presence of soliton solutions of this equation \cite{ZK}, and the inverse scattering technique developed for the KdV equation enabled the derivation of analytic solutions for given initial conditions with zeros at infinity \cite{GGKM}. Being the historic first among the integrable nonlinear evolution equations, the KdV equation attracted significant attention from both physicists and mathematicians \cite{Miura,Bullough,Miles,Lakshmanan}.\\\indent
Despite such extensive interest, until very recently, the KdV equation was thought to lack rogue wave solutions. This is true, but only if the wave described by the KdV equation is purely real. 
If we consider complex-valued solutions to the KdV equation, it is possible to derive rogue wave solutions \cite{Shallow}. Being the first work on this subject, the paper \cite{Shallow}, however, presented only selected (fixed-parameter) rogue wave solutions and did not reveal the large variety of possible features of this important class of solutions. In this work, we provide a more detailed mathematical treatment and derive families of rogue wave solutions with free parameters that determine a range of features. 
The presence of several parameters in our equations makes our approach much more powerful than in previous work \cite{Shallow}. 
Here, by use of a simple symmetry of the $n$-fold Darboux transformation, we show that rational solutions to the KdV equation can be substantially generalised to describe a much larger variety of rogue waves. In principle, depending on the choice of parameters involved, these rational solutions may be either singular or nonsingular. We also show that higher order rogue waves in the complex KdV equation can appear in multi-peak formations, in a similar way to the rogue waves of the NLS equation \cite{Kedziora,tri,triplet}.\\\indent
\section{The $n$-th Order Rational Solution for the Complex KdV Equation}
\begin{figure}[ht]
\includegraphics[scale=0.11]{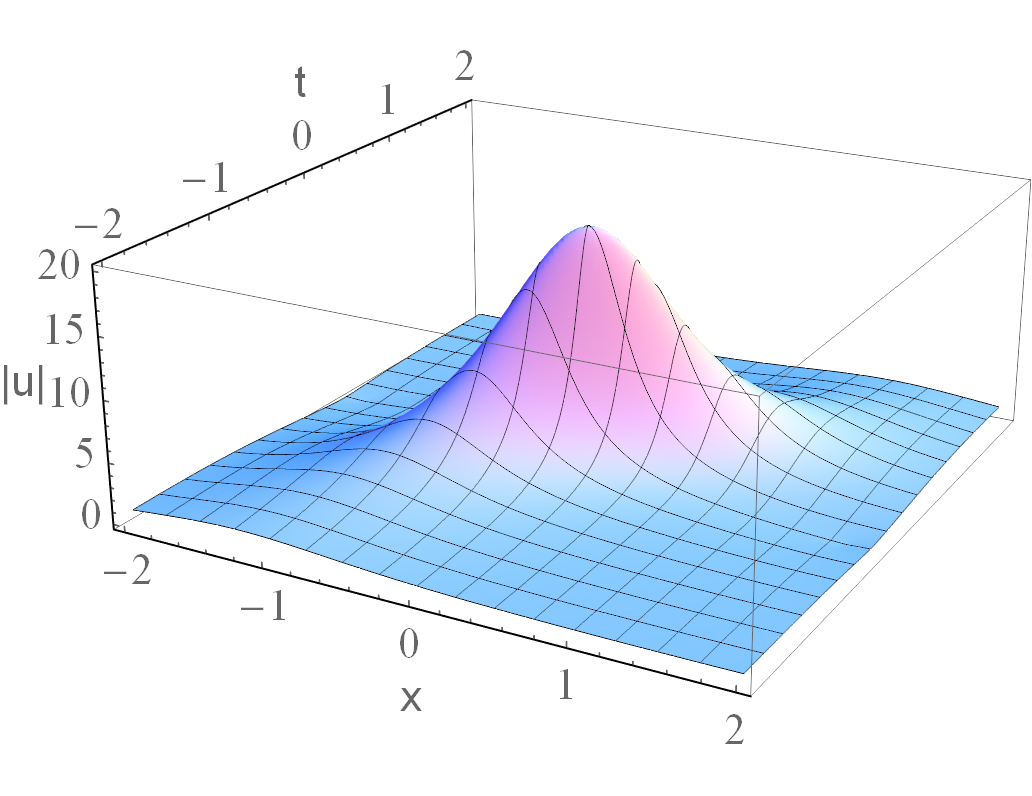}\includegraphics[scale=0.14]{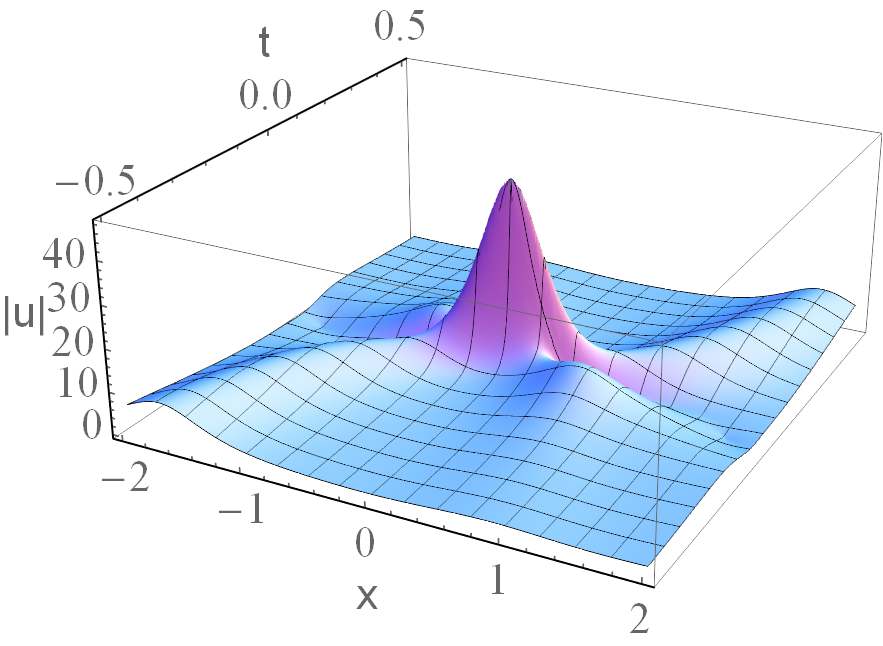}
\caption{\it (a) The plot of the second-order rogue wave (\ref{o2b0}) with $y=\frac32$ and $t_0=\frac12$. 
The maximal amplitude is $|u_2(0,0)| 
=20.0816.$ (b) The plot of the third-order rogue wave, defined by (\ref{sup}). Parameters are $y=\frac12,$ $t_0=-\frac{1}{24},$ and $g=-\frac65i$. The maximal amplitude is $|u_3(0,0)|=50$.  The background for both cases is $c=0.$}
\label{2nd}
\end{figure}

We will consider the complex KdV equation in the form
\begin{equation}
\label{kdv}
\frac{\partial u}{\partial t}-6u\frac{\partial u}{\partial z}+\frac{\partial^3u}{\partial z^3}=0.
\end{equation} 
where $z=x+iy$ is a complex variable, and $u=u(z,t)$ a complex function. Regardless of whether $u$ is real or complex, (\ref{kdv}) is also the condition of compatibility of the system
\begin{align} \label{2}
&	\frac{\partial\psi}{\partial t}=-4\frac{\partial^3\psi}{\partial z^3}+6u\frac{\partial\psi}{\partial z}+3\frac{\partial u}{\partial z}\psi,\\
&	-\frac{\partial^2\psi}{\partial z^2}+u\psi=\lambda\psi.  \label{3}
\end{align}
This equivalence has several consequences. One of the most important for our purposes is that the system (\ref{2}, \ref{3}) is Darboux covariant, giving us a dressing method to construct nontrivial solutions to (\ref{kdv}) from simple ones. Given an initial (seed) solution $u=u_0$ to the KdV equation, and $n$ linearly independent solutions $\psi_1,\dots,\psi_n$ to the associated linear system (\ref{2}, \ref{3}), with corresponding spectral parameters $\lambda=\lambda_1,\dots,\lambda=\lambda_n,$ the $n$-fold Darboux transformation of $u_0$ is given by \cite{Matveev}
\begin{equation}
\label{Darboux}
u_n=u_0-2\frac{\partial^2}{\partial z^2}\log W_{n+1},
\end{equation}
where  $W_n$ is the Wro\'nskian determinant of the functions $\psi_1,\dots,\psi_n$ with respect to $z$:
\begin{equation}\label{det}
W_n=\left|\begin{array}{cccc}
\psi_1 & \psi_2 & \dots & \psi_n\\
\partial_z\psi_1 & \partial_z\psi_2 & \dots & \partial_z\psi_n\\
\vdots & \vdots & \ddots & \vdots\\
\partial_z^{\;n-1}\psi_1 & \partial_z^{\;n-1}\psi_2 & \dots & \partial_z^{\;n-1}\psi_n
\end{array}\right|,\end{equation}
and $u=u_n$ will be another solution to the KdV equation (\ref{kdv}). To be more concise we omit writing explicitly the dependence of $W_n$ on the functions $\psi_1,\dots,\psi_n$.\\\indent
In order that the transformation (\ref{Darboux}) be non-trivial, the parameters $\lambda_k$ must be distinct. In order to obtain a Darboux transformation in the degenerate case $\lambda_k\to\lambda$ for all $k=1,2,\dots,n$, we define $\psi_1,\dots,\psi_n$ such that $\psi(z,t;\lambda_k)=\psi_k(z,t).$ Then, expanding the matrix element $\partial^{\;i}_z\psi_j$ as a Taylor series with respect to $\lambda_k,$ we have
\begin{equation}
\label{degdet}
\lim\limits_{\substack{\lambda_k\to\lambda\\
1\leqslant k\leqslant n}}W_n=\left|\begin{array}{cccc}
\psi & \partial_\lambda\psi & \dots & \partial_\lambda^{\;n-1}\psi\\
\partial_z\psi & \partial_\lambda\partial_z\psi & \dots & \partial_\lambda^{\;n-1}\partial_z\psi\\
\vdots & \vdots & \ddots & \vdots\\
\partial_z^{\;n-1}\psi & \partial_\lambda\partial_z^{\;n-1}\psi & \dots & \partial_\lambda^{\;n-1}\partial_z^{\;n-1}\psi
\end{array}\right|,\end{equation}
i.e. in the degenerate limit $W_n$ becomes the Wro\'nskian of the functions $\psi,\partial_\lambda\psi,\dots,\partial_\lambda^{\;n-1}\psi.$\\\indent
So if we take, for example, the simple constant seed solution $u_0(z,t)=c,$ we can take as linearly independent solutions to the system (\ref{2}, \ref{3}) the functions
\begin{equation}\label{sol-n}
\psi_k(z,t)=\cosh\omega_k\{z+2(3c-2\omega_k^{\;2})t\}
\end{equation}
where
$\omega_k=\sqrt{c-\lambda_k}~\text{for}~\lambda_k\neq c$.
In (\ref{sol-n}), each eigenvalue $\lambda_k$ is distinct. \\\indent
We will also note here that due to the translational invariance
$u(z,t)\mapsto u(z-z_0,t-t_0)$
of the KdV equation (\ref{kdv}), we can introduce a second constant, via $t\mapsto t-t_0.$ This will not affect the choice of $\psi$ in any substantial way, but this will be relevant later, so we allow for arbitrary shifts in $z$ and $t$.
For simplicity's sake, we set a background $c=0,$ and the function $\psi$ in (\ref{sol-n}) becomes 
\begin{equation}
\label{wg}
\psi(z,t;\lambda_k)=\cos\sqrt{\lambda_k}\{z+4\lambda_k(t-it_0)\},
\end{equation}
from which we get
\begin{align*}W_2(z,t)=-\tfrac12z-6\lambda(t-it_0)-\frac{\sin2\sqrt{\lambda}\{z+4\lambda(t-it_0)\}}{4\sqrt{\lambda}},\end{align*}
with $W_2(z,t)\to-z$ as $\lambda\to0.$\\\indent
The imaginary part of $z$ ensures that the Wro\'nskian $W_2(z,t)$ has no zeros in $z$ and $t$ except for the origin, and in the limit as $\lambda\to c,$ in this case as $\lambda\to0,$ the Wro\'nskian becomes a polynomial in $x$ and $t$. The corresponding degenerate solution $u_1$ to the complex KdV equation will thus always be a non-singular rational function as $\lambda\to0$ if $y$ is chosen appropriately. In the simplest case, $u_1(x,t)$ becomes 
\[u_1(z,t)=\frac{2}{z^2}=\frac{2}{(x+iy)^2}.\]
The $3\times3$ Wro\'nskian becomes in the limit 
\[\lim\limits_{\lambda\to0}W_3(x,t)=-\tfrac23(x+iy_0)^3-8(t-it_0).\]\indent
For better clarity, we will write $K_n$ for the limit of the Wro\'nskian $W_{n+1}$ as $\lambda\to0,$ i.e.
\begin{equation}
	K_n(x+iy_0,t-it_0)=\lim\limits_{\lambda\to0}W_{n+1},
\end{equation}
so that in general, the $n$-th order rational solution of the KdV equation is given by
 \begin{equation}
 u_n(x,t)=-2\frac{\partial^2}{\partial x^2}\log K_n(x,t).
 \end{equation}
 \indent
 Singularities do not appear in all parts of the complex plane. If we choose the parameters of the solution such that the system\[\Re\{K_n(x+iy,t-it_0)\}=0, ~\Im\{K_n(x+iy,t-it_0)\}=0\]
has no solution in real values of $x$ and $t.$ To find a nonsingular second-order solution, we observe that if $t_0$ is strictly real, then
\[\Im\{K_2(x+iy,t-it_0)\}=2(x^2y-\tfrac13y^{\;2}+4t_0),\]
and this is a quadratic in $x$ with no real zero in $x$ in the region 
\[\frac{y^{\;3}+12t_0}{y}<0.\]
In this region of the complex plane, the solutions will not have any singularities for any values of $x.$\\\indent
The constant $t_0$ allows us to avoid any real zeros in the denominator of the second-order solution, since if $t_0=0,$ the equation $K_2(x+iy,t-it_0)=0$ always has roots in real values of $x$ and $t$ for any choice of $y.$ \\\indent
The second-order rational solution of the complex KdV equation is
\begin{equation}
\label{o2b0}
u_2(z,t)=6z\frac{z^3-24(t-it_0)}{\{z^3+12(t-it_0)\}^2}.
\end{equation}
The plot of this function for fixed parameters $y$ and $t_0$ is shown in Fig. \ref{2nd}. 
With these restrictions on $y$, this has the form of a rogue wave with the maximal amplitude at the origin, being given in terms of $y$ and $t_0$ by
\[|u_2(0,0)|=6\frac{|y^{\;4}+24yt_0|}{(y^{\;3}-12t_0)^2}.\]

It is a straightforward exercise to write rational solutions of any order $n$. To give a few more examples, the Wro\'nskians $K_n$ as $\lambda\to0$ for the third to fifth-order solutions are
\begin{eqnarray*}
K_3(z,t) &=&\frac{4}{15}(z^6+60z^3t-720t^2), 
 \\
K_4(z,t)&=&\frac{32}{515}(z^{10}+180z^7t+302400zt^3),  
	 \\
K_5(z,t)&=&\frac{256}{33075}(z^{15}+420z^{12}t+25200z^9t^2+\\
	&+&2116800z^6t^3-254016000z^3t^4-
	\\
	&-&1524096000t^5).
\end{eqnarray*} up to translations $z\mapsto z+z_0,~t\mapsto t-it_0.$\\\indent
Since $K_n(z,t)$ is an $(n+1)\times(n+1)$ determinant, the explicit formulae quickly become more cumbersome with increasing $n$, although the description in terms of the determinant holds for all $n$. The second-order rogue wave in Fig. \ref{2nd}(a) is the simplest one among the hierarchy of KdV rogue waves. It has a single maximum, and smoothly growing and decaying fronts.\\\indent
The symmetry
\begin{equation}
\label{galilean}
	u(z,t)\mapsto c+u(z+6ct,t)
\end{equation}
of the KdV equation allows us to chose the background $c$ arbitrarily, but at the expense of a travelling velocity of $6c.$ 
The two lowest order solutions with this adjustment become
\begin{align*}
	u_1(z,t)&=c-\frac{2}{(z+6ct)^2},\\
	u_2(z,t)&=c-2\frac{\partial^2}{\partial z^2}\log\{(z+6ct)^3+12(t-it_0)\}.
\end{align*}
These are generalisations of previously obtained rogue wave solutions. Namely, when  $c=-1,$ $y=-\frac12$ and $t_0=\frac{1}{24}$, the above solutions coincide with those derived in \cite{Shallow} by the complex Miura transformation. 
\section*{Multi-peak solutions}
\begin{figure}[ht]
	\includegraphics[scale=0.14]{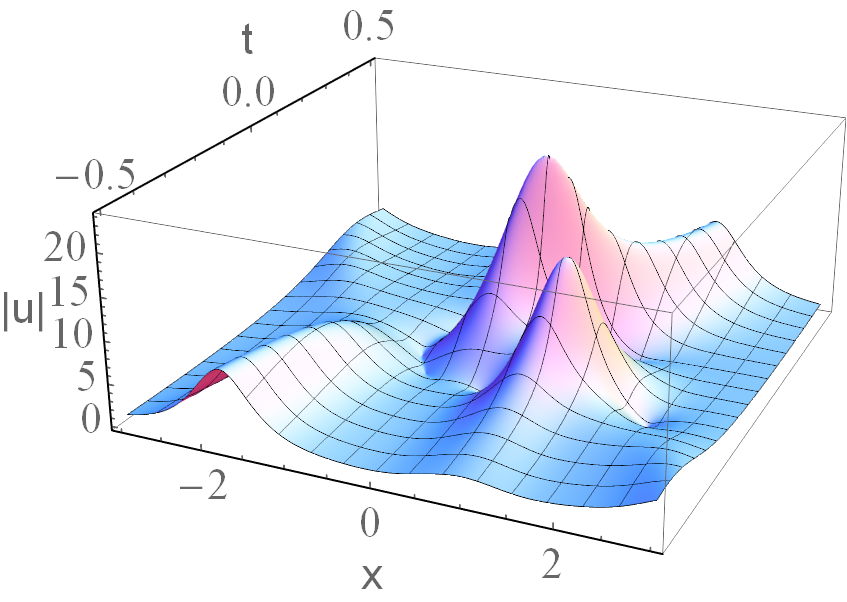}\includegraphics[scale=0.14]{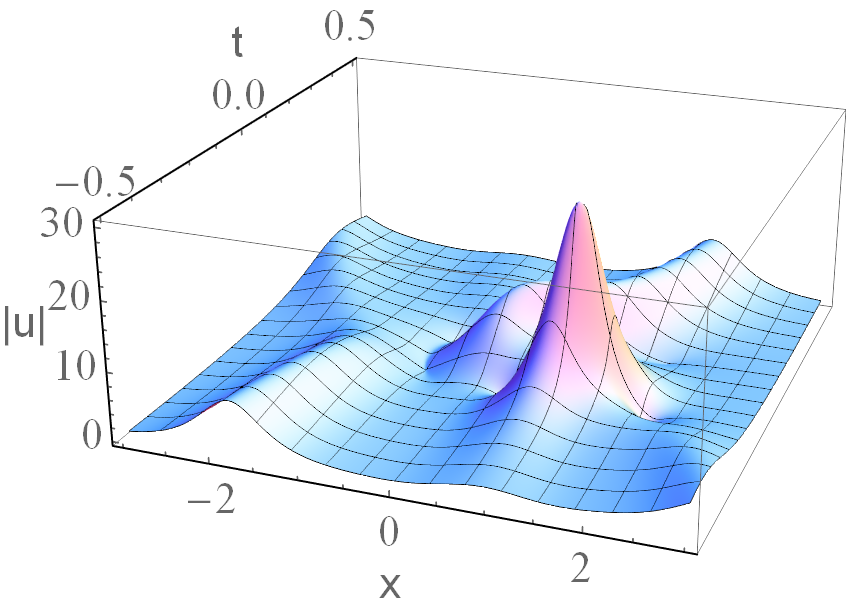}
	\caption{\it  Two plots of the third-order rogue wave. (a) Parameters are the same as in Fig. \ref{2nd}(b) except now $g=-5-\frac65i.$ A second peak has grown and the larger peak has decreased in total amplitude and moved out from the origin. (b) Parameters are the same as in (a) except now $t_0=-\frac{1}{40}.$ One of the peaks has faded and the other peak has gained more amplitude.}
	\label{3rdsep}
	\end{figure}

The higher-order solutions of the hierarchy are more complicated. Moreover, the usual expressions for them are always singular and may not describe physical situations. In order to obtain solutions which can be nonsingular, we have to go beyond the standard dressing technique. \\\indent
Solutions of order $n\geqslant3$ can be further generalised by making use of an empirically found symmetry which allows us to replace the functions $K_n$ with linear combinations of $K_n$ and $K_{n-2}:$ If
\begin{equation}
\label{sup}
	V_n(z,t;g)=K_{n}(z,t)-gK_{n-2}(z,t),~n=3,4,\dots,
\end{equation}
where $g$ is an arbitrary constant, then 
\begin{equation}
u=u_n(z,t;g)=-2\frac{\partial^2}{\partial z^2}\log V_n(z,t;g)~(n\geqslant3)
\end{equation}
is also a solution of the KdV equation (\ref{kdv}). This symmetry can be verified by direct substitution as we have done for all cases found here, i.e. up to $n=5,$ and we conjecture it holds in general for all $n$.\\\indent
If we were to extend this symmetry to the $n=2$ case, then it would make sense to identify $K_0$ as simply a constant, since this would generate the solution $u=0.$ Then (\ref{sup}) for $n=2$ would just be introducing a complex additive constant. If $\Re(g)=0,$ it is identical to the substitution $t\mapsto t-it_0.$\\\indent
When $n=3,$ we recover a third-order rational solution,
\begin{align}
\label{3g}
u_3(z,t;g)&=\frac{P_3(z,t-it_0;g)}{\{V_3(z,t-it_0;g)\}^2},\\
P_3(z,t;g)&=450g^2+2160gz^5+8294400zt^3+\nonumber\\
&~~~+1036800z^4t^2+192z^{10},
\end{align} 
up to the symmetry (\ref{galilean}).
The parameters $t_0$, $g$ and $c$ here allow us to control the shape of the rogue wave. One example of this solution for background $c=0$, with $y=\frac12,$ $t_0=-\frac{1}{24}$ and $g=-\frac65i$ is shown in Fig. \ref{2nd}(b). This same choice of parameters with background $c=-1$ again reduces to the particular third-order rogue wave solution given in \cite{Shallow}.
This, like (\ref{o2b0}), has a single central peak at the origin, but now there are two sets of tails, resembling a two-soliton collision.\\\indent 
When the parameter $g$ is purely imaginary, the rogue wave is dominated by its central peak. The imaginary part of $g$ affects the relative heights of the peaks and the tails.  When $\Im(g)$ becomes large, the peaks reduce in size relative to the tails, and for sufficiently large values, the peaks may be even smaller than the tails. On the other hand, the real part of $g$ causes splitting of the lower order components, so that they do not directly collide at the origin. Instead, with $\Re(g)\neq0,$ we see growth of multiple peaks. In Fig. \ref{3rdsep}(a), the central peak splits into two smaller ones, their locations and amplitudes depending on $g$ and $t_0.$ Another example is shown in Fig. \ref{3rdsep}(b), in which we see that the effect of the parameter $t_0,$ and position $y$ on the imaginary axis, can be to transfer amplitude from one peak to another.

\begin{figure}[ht]
	\includegraphics[scale=0.14]{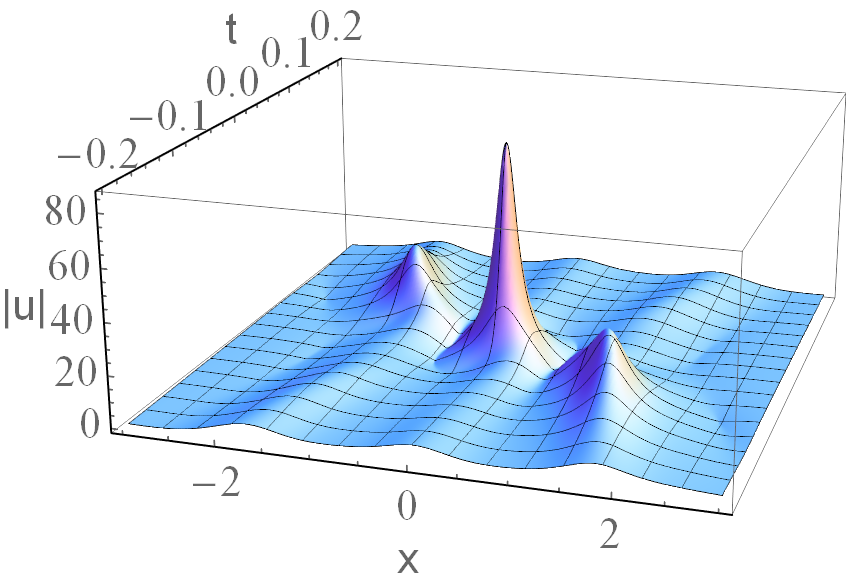}\includegraphics[scale=0.14]{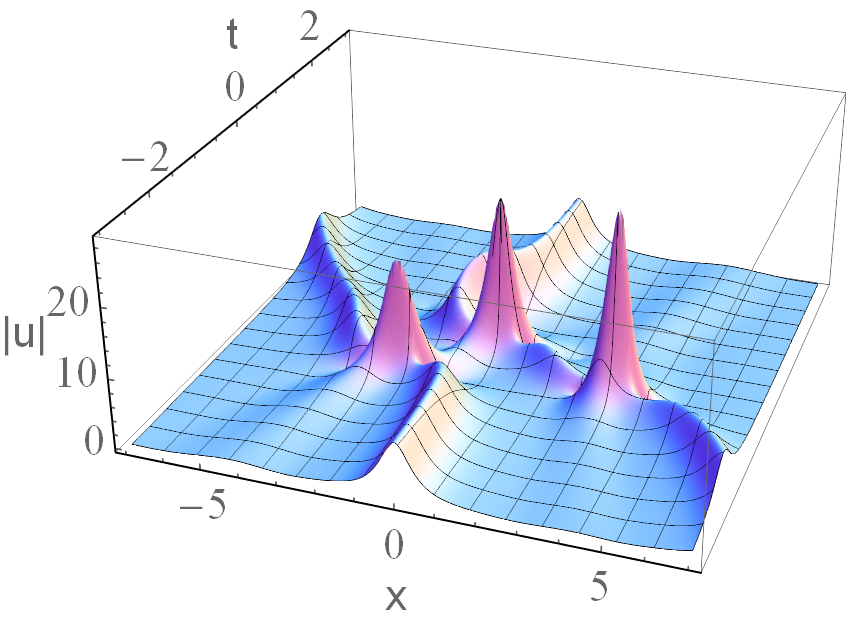}
	\caption{\it Two plots of the fourth-order rogue wave. (a) Parameters are $y=-\frac12,$ $t_0=\frac{1}{24},$ $g=-6i$ and $c=0$. The maximal velocity is $|u_4(0,0)|=82$. (b) Parameters are the same except now $g=1888-12i.$ With $g$ having nonzero real part, the main peak has decreased in amplitude and the smaller peaks have grown and moved away from the origin.}
	\label{4th}
\end{figure}

The general fourth-order rational solution in explicit form is given by
\begin{align}
\label{4g}
&~~~~~~~  u_4(z,t;g) =\frac{P_4(z,t-it_0;g)}{\{K_4(z,t)-gK_2(z,t)\}^2}, \\
&P_4(z,t;g)= 50 \{32z^9+4032z^6t+967680t^3+105gz^2 \}^2-  \nonumber \\
&~~~-60z\{35g+48z^4(z^3+84t)\}\{16z^{10}+  \nonumber \\
&~~~+2880z^7t+4838400zt^3+175g(z^3+12t)\}. 
\end{align}
The explicit form of the fifth-order rational solution is
\begin{align}
&u_5(z,t;g)=\frac{P_5(z,t-it_0;g)}{\{K_5(z,t)-gK_3(z,t)\}^2},\\
&P_5(z,t;g)=60z\{972405g^2(43200t^3+5400t^2z^3+z^9)+\nonumber\\
&~~~+2048(-154857907814400000t^9+\nonumber\\
&~~~+19357238476800000z^3t^8+184354652160000z^6t^7+\nonumber\\
&~~~+17667320832000z^9t^6+512096256000z^{12}t^5+\nonumber\\
&~~~+1447891200z^{15}t^4+120960z^{21}t^2 + 504z^{24}t +z^{27}) \}-\nonumber\\
&~~~-282240g(9144576000t^6-21772800z^6t^4-\nonumber\\
&~~~-483840z^9t^3+22680z^{12}t^2+252z^{15}t+z^{18}).
\end{align}
Higher order rational solutions can also be written in similar form, but quickly exceed reasonable limits of presentability.\\\indent
Two examples of the fourth-order solution for given sets of parameters and zero background $c$ are shown in Fig. \ref{4th}. The profile of this solution can again take a multiplicity of forms.  When the parameter $g$ is purely imaginary, as in Fig. \ref{4th}(a), most of the rogue wave amplitude is concentrated in the central peak, although two small, symmetrically located side peaks are also present. The maximal amplitude of the central peak here is $82$. 
\\\indent An example of the fourth order rogue wave for $g$ with nonzero real part is shown in  Fig. \ref{4th}(b). Again, the real part of $g$ causes multiple peaks to grow. There we have three distinct large peaks but of smaller amplitudes, each roughly 20 to 30. Their relative locations and values of velocity  are again determined by $g$ and $t_0$.

This extreme localisation shown in all graphs is the characteristic feature of rogue waves \cite{WANDT}. 
\section{Conclusion}
The crucial step taken in our work is the generalisation (\ref{sup}). Despite being as simple as a linear superposition law for the Wro\'nskians, it has important nontrivial consequences for the whole family of rational solutions, allowing them to be nonsingular i.e. physically relevant rogue waves. It also adds the complex parameter $g$ that is essential in the higher order rational solutions for removing singularities and for splitting the higher-order rogue wave into its fundamental components. When the real part of $g$ is zero, then the rogue wave has the highest peak at the origin and smaller local maxima around it, as shown in Fig. \ref{4th}(a). On the other hand, when $g$ has nonzero real part, the central large peak decreases and smaller side peaks grow while separating from each other. This type of splitting of higher order rogue waves into multiplet structures has also been observed in the case of NLS rogue waves \cite{Kedziora,tri,triplet} and their extensions \cite{2b}. However, the splitting of the higher-order rogue waves of the complex KdV equation is more complicated. The complete classification of all forms of rogue waves here remains open for investigation.\\\indent
Lastly, we point out that complex solutions of the KdV equation are also applicable to unidirectional crystal growth \cite{crystals} and complex KdV-like equations serve to model dust-acoustic waves in magnetoplasmas \cite{misra1}. The KdV hierarchy itself also finds applications in modern theories of quantum gravity \cite{ijp}.\\\indent
As such, these new solutions presented in this work, previously not thought to exist, may find much wider use in various areas of physics.

\bibliographystyle{apsrev4-1}

\begin{thebibliography}{99}

\bibitem{Kharif} C. Kharif, E. Pelinovsky, A. Slunyaev,  Rogue Waves in the Ocean. (Springer, Berlin-Heidelberg, 2009).
\bibitem{Osborne} A. Osborne, Nonlinear Ocean Waves and the Inverse Scattering
Transform (Elsevier, Amsterdam, 2010).
\bibitem{Grimshaw} R. Grimshaw, E. Pelinovsky, T. Taipova, and A. Sergeeva,
Eur. Phys. J. Spec. Top. {\bf 185}, 195 (2010).
\bibitem{Alford} M. H. Alford, Nature {\bf 521}, 65 (2015).
\bibitem{Solli} D. R. Solli, C. Ropers, P. Koonath \& B. Jalali, Optical rogue waves, Nature {\bf 450}, 1054 (2007).
\bibitem{Dudley} J. M. Dudley, {\it et al}, Optics Express, {\bf 17}, 21497 (2009).
\bibitem{Chekhova} M. Manceau, K. Yu. Spasibko, G. Leuchs, R. Filip, M. V. Chekhova, 
Phys. Rev. Lett., {\bf 123}, 123606 (2019). 
\bibitem{Bayindir} C. Bayindir and M. Arik, Rogue quantum gravitational waves, arxiv.org/abs/1908.02601v1 (2019).
\bibitem{Chabchoub}  A. Chabchoub, N. P. Hoffmann, and N. Akhmediev, 
Phys. Rev. Lett., {\bf 106}, 204502 (2011). 
\bibitem{Schiek} R. Schiek, and F. Baronio, 
Phys. Rev. Res., (2019), Manuscript LE 17126 accepted for publication.
\bibitem{Nikolkina} I. Nikolkina and I. Didenkulova,
Nat. Hazards Earth Syst. Sci. {\bf 11}, 2913 -- 2924, (2011).
\bibitem{Baronio} F. Baronio, A. Degasperis, M. Conforti, S. Wabnitz, 
Phys. Rev. Lett. {\bf 109}, 044102 (2012).
\bibitem{Gaillard} P. Gaillard, J. Phys. A {\bf 44}, 435204 (2011).
\bibitem{Trulsen} K. Trulsen, K. B. Dysthe,  Wave motion, {\bf 24}, 281 (1996).
\bibitem{Dubard}
P. Dubard, V. B. Matveev, Nat. Hazards Earth. Syst. Sci. {\bf 11}, 667 (2011).
\bibitem{Kodama} Y. Kodama, J. Phys. A: Math. Theor. {\bf 43}, 434004 (2010).
\bibitem{Ohta}  Y. Ohta and J. Yang, Phys. Rev. E 86, 036604,(2012).
\bibitem{Zhang}  Y. Zhang, D. Qiu, D. Mihalache, and J. He, Chaos {\bf 28}, 103108 (2018).
\bibitem{Shrira} V. I. Shrira, V. Geogjaev, J. Eng. Math. {\bf 67}, 11, (2010).
\bibitem{Onorato} M. Onorato, S. Residori, U. Bortolozzo, A. Montina, F. T. Arecchie,  
Sci. Rep. {\bf 528}, 47-89 (2013).
\bibitem{Korteweg} D. J. Korteweg \& G. De Vries, Phil. Mag. {\bf 39}, 422 (1895).
\bibitem{Boussinesq}
J. Boussinesq, Acad. Sci. Inst. Nat. France, {\bf XXIII}, 1 -- 680 (1877).
\bibitem{ZK} N. J. Zabusky and M. D. Kruskal, Phys. Rev. Lett. \textbf{15}, 240 (1965)
\bibitem{GGKM} C. S. Gardner, J. M. Greene, M. D. Kruskal, R. M. Miura, Phys. Rev. Lett. {\bf 19}, 1095 -- 1097 (1967).
\bibitem{Miura} R. M. Miura, SIAM Review, {\bf 18}, No. 3, 412 -- 459,  (1976). 
\bibitem{Bullough}
R.  K. Bullough, and  P.  J. Caudrey, Acta Appl. Math., {\bf 39}, 193 -- 228, (1995).
\bibitem{Miles} J. W. Miles, J. Fluid Mech., {\bf 106}, 131--147, (1981).
\bibitem{Lakshmanan}  M. Lakshmanan, S. Rajasekar,  Advanced Texts in Physics, (Springer, Berlin, Heidelberg, 2003)
\bibitem{Shallow} A. Ankiewicz, M. Bokaeeyan, and N. Akhmediev, 
Phys. Rev. E \textbf{99}, 050201(R) (2019).

\bibitem{Kedziora} D. J. Kedziora,  A. Ankiewicz, and N. Akhmediev, 
Phys. Rev. E {\bf 88}, 013207 (2013).

\bibitem{tri} A. Ankiewicz and N. Akhmediev, Rom. Rep. Phys. \textbf{69}, 104 (2017)
\bibitem{triplet} A. Ankiewicz, D. J. Kedziora, N. Akhmediev,  Phys. Lett. A {\bf 375}, 2782 (2011).
\bibitem{Matveev} V. B. Matveev, M. A. Salle, Darboux Transformations and Solitons, (Springer, Berlin-Heidelberg, 1991)  
\bibitem{WANDT} N. Akhmediev, A. Ankiewicz and M. Taki, 
Phys. Lett. A \textbf{373}, 675 (2009).


\bibitem{2b} M. Crabb \& N. Akhmediev, Nonlin. Dyn. \textbf{98}, 245 (2019).
\bibitem{crystals} M. Kerszberg, Phys. Lett. \textbf{105}A, 4, 5 (1984) 
\bibitem{misra1} A. Misra, Appl. Math. Comp. \textbf{256}, 386--374 (2015)
\bibitem{ijp} R. Iyer, C. V. Johnson, J. S. Pennington, arXiv:1002.1120v1 [hep-th] 5 Feb 2010

\end{thebibliography}

\end{document}